\documentclass[prb,superscriptaddress,amsmath,amssymb,twocolumn]{revtex4}
\usepackage{graphicx}
\usepackage{color}

\begin{document}

\title{
Competing supersolids of Bose-Bose mixtures in a triangular lattice
}

\author{Fabien Trousselet}
\affiliation{Institut N\'eel, Universit\'e Grenoble Alpes and CNRS, F-38042 Grenoble, France}
\author{Pamela Rueda-Fonseca}
\affiliation{CEA, INAC-SP2M,  F-38054 Grenoble, France }
\author{Arnaud Ralko$^{1}$}
\date{\today}


\begin{abstract}
We study the ground state properties of a frustrated two-species mixture of
hard-core bosons on a triangular lattice, as a function of tunable amplitudes
for tunnelling and interactions. 
By combining three different methods, a self-consistent cluster mean-field,
exact diagonalizations and effective theories, we unravel a very rich and
complex phase diagram.
More specifically, we discuss the existence of three original mixture
supersolids: (i) a commensurate with frozen densities and supersolidity in spin
degrees of freedom, in a regime of strong interspecies interactions; and (ii)
when this interaction is weaker, two mutually competing incommensurate
supersolids.  Finally, we show how these phases can be stabilized by a quantum
fluctuation enhancement of peculiar insulating parent states.

\end{abstract}

\maketitle

\section{\label{IN}Introduction}

In nowadays condensed matter physics, common fascinating collective behaviors
and novel quantum phases are reported in various domains of physics, more
specifically in bosonic systems encountered in quantum magnetism, ultracold
atoms on optical lattices and strongly correlated materials.
Thanks to their versatility, an important amount of exotic phases has been
reported in the literature these last years, both experimental and theoretical,
such as different types of superfluids \cite{Jaksch98}, insulators
\cite{albuquerque,Ralko01}, Bose metals \cite{bose1, bose2, bose3} or
supersolid phases \cite{Ralko02,Kim4,PS05,Capo10,WesAl,CMF2}.
In the latter, the system enters a phase combining crystalline order
and superfluidity, and typically arising from the quantum melting of a Mott
insulator; this phenomenon is at the origin of intense scientific activities
and debates;
Experimentally, indications for supersolidity were supposed to be found in
$^4$He \cite{Kim4}. However, necessary conditions for continuous symmetry
breaking questioned this interpretation \cite{PS05} and recent experiments
have clearly ruled out this scenario \cite{Kim12}.

Meanwhile, supersolids are easier achieved on lattice systems; whereas on a
square geometry with nearest neighbor interactions a soft-core description is
required \cite{Seng5}, they can also be found in hard-core bosonic models when
frustration is induced by either further neighbor interactions \cite{Capo10}
and/or lattice geometry, {\it e.g.} triangular \cite{Pol10,WesAl,CMF2}.
An other interesting direction to stabilize supersolidity is to increase the number
of degrees of freedom in frustrated systems. In that respect, bosonic mixtures
with several species of bosons, usually encountered in optical lattices
\cite{optri,opkag}, either heteronuclear \cite{Cat8} or homonuclear
\cite{Gad10}, as well as  bilayer systems with interactions but no hopping
between layers \cite{dipatt} are very promising.
Such systems allow for an even broader variety of quantum phases than their
single-species counterparts, depending on the intra- and inter-species
interactions, and on the dimension and lattice connectivity. 

Theoretically, compared to an already rich literature on mixtures in 1D
\cite{Buon8} only few works have focused on 2D cases for instance on square
\cite{2DUab} and triangular \cite{mixtJ,mixDMFT,BFtri} lattices. 
In all those systems, when contact and dipolar interactions
\cite{dipol,dipatt,CaS} are taken independently, interaction-induced
insulators, {\it e.g.} density-homogeneous or density wave \cite{CRFMF,MRF} are
favored. Hence,  their competition can, along with quantum fluctuations,
trigger various unconventional phases \cite{dipatt,CaS,WesAl,Tiel}, especially
in systems with kinetic or interaction frustration.

\begin{figure}[h!]
\begin{center}
\includegraphics[width=0.30\textwidth]{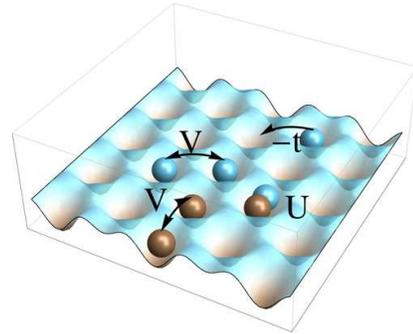}
\caption{\label{mopha}
Triangular lattice hosting bosons of two species labelled $a$ (blue) and
$b$ (brown).
the bosons can hope from two neighboring sites with the energy $-t$, and
interact via an intra-species interaction $V$ and a point contact inter-species
interaction $U$.
}
\end{center}
\end{figure}

In this paper, we provide a theoretical study of a two-species bosonic mixture
on a triangular lattice as depicted in Fig.~\ref{mopha}.
We aim to focus on the most simple model with competing interactions in
frustrated geometry in order to study possible mechanisms for stabilizing
exotic phases.
The richness of the phase diagram of a related one-species model with hard core
constraint \cite{CMF2} allows us to expect even more exotic physics for a
mixture of two mutually interacting species.
Even though such interactions can be encountered in cold atom systems
\cite{WesAl,CMF2,dipatt} with polarized dipoles \cite{dipol,lutchyn,armstrong},
the present work describes the maximally frustrating cases for which all
interactions are repulsive.
We address the following issues: (i) how on-site and nearest neighbor
interactions compete and (ii) how lattice frustration impacts on the
stabilization of non-conventional phases of such a mixture.

To achieve these goals, we use mainly a {\it cluster mean field} theory (CMFT) and
exact diagonalizations (ED) on periodic clusters; both methods are detailed, along with 
the model we consider, in Section \ref{MM}.
Note that preliminary Quantum Monte-Carlo (QMC) simulations have been performed
to support our findings (see text).
We present in Section \ref{PD} an overview of the phase diagram obtained this way, 
before focusing on the most interesting phases. First, we describe in Section \ref{CS} some 
commensurate spin-like phases. These are characterized with means of perturbative approaches; 
they have frozen densities and, for one of them, spin-like supersolidity. Next, we analyze in 
Section \ref{IS} incommensurate phases found in this study: these include two original 
two-species supersolid (SS) phases, which belong to our main findings. Eventually we address in 
Section \ref{PT} the nature of phase transitions involving these peculiar phases, before some 
concluding remarks in Section \ref{CC}.

\section{\label{MM} Model and method} We study a two-species (spin) extended Bose-Hubbard
model on a triangular lattice:
\begin{eqnarray}
H_{\alpha}&=&  \sum_{\langle i,j\rangle}
\left[-t_\alpha (b^\dagger_{i \alpha}b_{j \alpha} + h.c.) + V_\alpha
n_{i \alpha}n_{j \alpha} \right] -  \mu_\alpha \sum_{i} n_{i \alpha}
\nonumber\\
H_{ab} &=& \sum_{\alpha} H_\alpha + U \sum_i n_{i a}n_{i b}.
\label{ham}
\end{eqnarray}
$H_\alpha$ is the one-species Hamiltonian ($\alpha =a,b$), with $t_\alpha$ and
$V_\alpha$ respectively the nearest neighbor hopping and interaction
amplitudes; $\mu_\alpha$ is the chemical potential for bosons of species
$\alpha$ created by operators $b_{i\alpha}^\dagger$ at site $i$. In this work,
we focus on the limit of hard core bosons ($U_{\alpha,\alpha} \gg |t_\alpha|,
V_\alpha, \mu_\alpha$) with an implicit onsite intra-species repulsion $
U_{\alpha \alpha}  \sum_{i} n_{i \alpha} (n_{i \alpha} -1 )/2$. 
Together with the repulsive interspecies coupling $U$, this allows us to maximize
the effects of frustration.
We have studied the more general case
in function of independent $\mu_a$ and $\mu_b$ and found that the richest
physics was found in the symmetric case $\mu_a=\mu_b$, including the novel
supersolid regimes.
Hence, when the system is $(a,b)$-symmetric ($t_\alpha=t$, $V_\alpha=V$ and
$\mu_\alpha=\mu$), as considered in this work since the most original two-spin
phases arise there, the particle-hole transformation
$b'_{i\alpha}=b^\dagger_{i\alpha}$ allows for a mapping between $\mu^*>0$ and
$\mu^*<0$, with $\mu^*=\mu-3V-U/2$ a rescaled chemical potential.
 Note that, such a transformation,  if restricted to a single species and for
$\mu^* = 0$ amounts to changing the sign of $U$. This $U<0$ case is of interest
since the corresponding terms mimic qualitatively the effects of interlayer
short-range interactions in dipolar cold atom bilayers \cite{dipatt}.

\begin{figure}[!ht]
\begin{center}
\includegraphics[width=0.25\textwidth,clip]{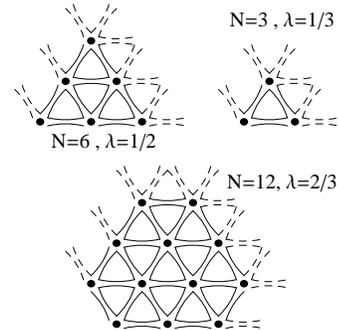}
\caption{ \label{clusters} Clusters considered in our CMFT analysis with $N=3$,
$6$ and $12$ sites. Internal and external bonds correspond respectively to
continuous and dashed lines. For each cluster, the cluster scaling parameter $\lambda$ is given.
}
\end{center}
\end{figure}

We compute the ground states of $H_{ab}$ on periodic clusters with up to $N=12$
sites (see Fig.\ref{clusters}), using either CMFT (if not precised) or ED
methods.
Note that the former method has been employed successfully in many one-species 
bosonic systems on various lattices such as
triangular \cite{CMF2,CMF3,CRFMF}, pentagonal \cite{pent} and hexagonal
\cite{albuquerque}. 
It is thus expected to be also very efficient in the present model.

In the former case, the  $N_i$ ($N_e$) internal (external) bonds are treated
exactly (at the mean field level) \cite{CMF1,CMF2,CMF3} and correlations are
better taken into account as $N$ increases.  As in Ref.~\cite{CMF3}, we thus
define the scaling parameter $\lambda = N_i / (3N)$ which quantifies finite
boundary effects; the Thermodynamic Limit (TL) is achieved for $\lambda \to 1$
(infinite lattice).
The mean-field parameters determined self-consistently are the densities
$\bar{n}_{i\alpha}= \langle n_{i\alpha} \rangle$ and the superfluid fractions
(SF) $\phi_{i\alpha}=\langle b_{i\alpha} \rangle$.
In addition, to evidence 3-fold symmetry breaking,  we define the order
parameter $M_\alpha = | \bar{n}_\alpha(k) |$, the diagonal $S_{ab}^{d} = |
\langle (n_a-n_b)(-k) (n_a-n_b)(k) \rangle |$ and off-diagonal $S_{ab}^{od} = |
\langle \sum_{i,j} b_{ia} b_{ib}^\dagger b_{jb} b_{ja}^\dagger  \rangle | / N^2
$ correlation functions. We used the Fourier transform $n_\alpha(k)
=\frac{1}{N} \sum_s n_{s \alpha} e^{i k\cdot r_s}$ at point $k = (4 \pi /3,
0)$, corner of the Brillouin zone.

\begin{figure}[ht]
\begin{center}
\begin{minipage}{0.33\textwidth}
\includegraphics[width=1.00\textwidth]{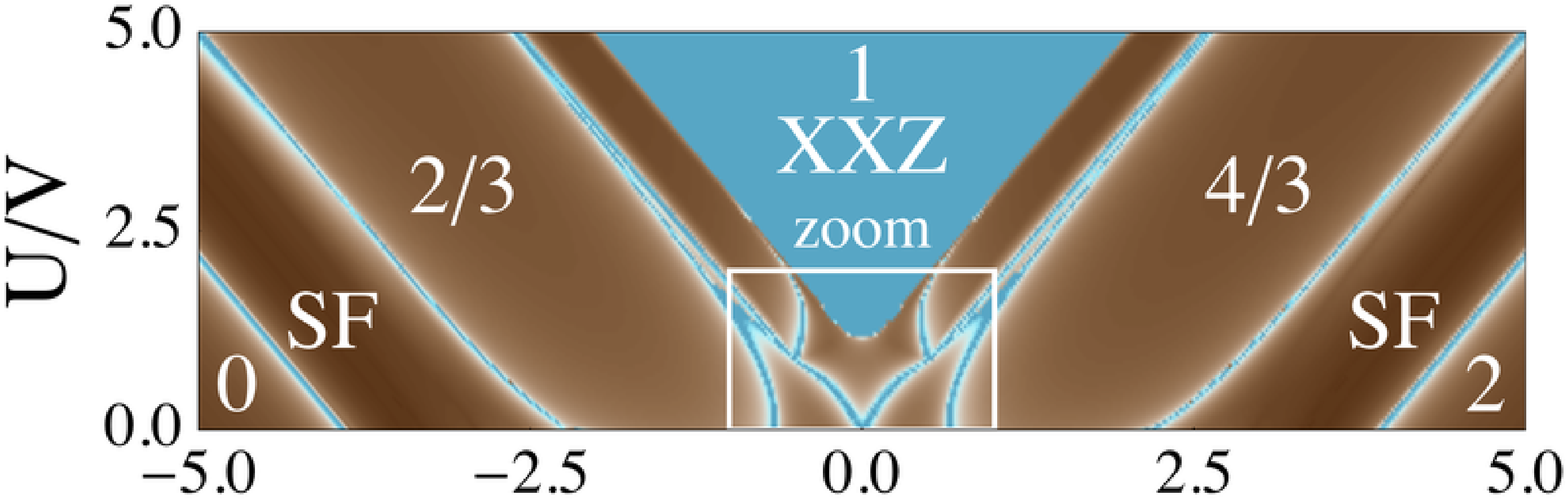}
\end{minipage}\\
\begin{minipage}{0.33\textwidth}
\includegraphics[width=1.00\textwidth]{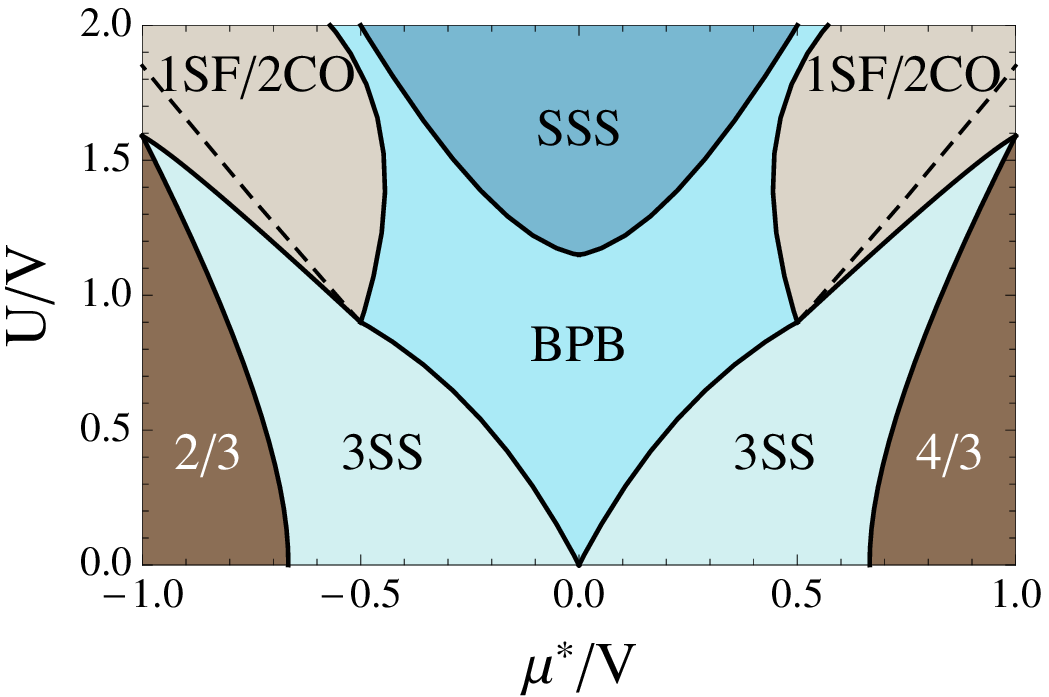}
\end{minipage}
\caption{\label{diag3t015} 
Phase diagrams as obtained from CMFT calculations on a $N=3$ cluster at
$t=0.15$ and $V=1$. Top: wide range of $\mu^*$ showing two-spin 
superfluids, 2/3 and 4/3  plateaus, XXZ physics domain and original collective
two-spin physics  (white rectangle).
Bottom: zoom of the white rectangle and all new two-spin phases (see text
for acronyms). Solid (dashed) lines are phase boundaries subsisting
(vanishing) in the Thermodynamic Limit (TL).
}
\end{center}
\end{figure}

\section{\label{PD} Overview of the phase diagram} 
Let us first focus on the $(\mu^*,U)$
phase diagram in which two-species phases - detailed along the paper -
emerge, illustrated here for $t/V=0.15$.
For these parameters, the one-species Hamiltonian $H_\alpha$ is known to present
a rich phase diagram with either empty/full, homogeneous superfluid,
$\sqrt{3}\times \sqrt{3}$ solid at density $1/3$ or $2/3$, or supersolid phases
\cite{WesAl}. 
When $U$ is switched on, the correlations between the two species increase, and
the resulting GS can be either a product of two one-species phases or a
species-entangled state as depicted in the phase diagram shown in
Fig.~\ref{diag3t015}.
(i) As $\mu^* < 0$ increases, the first non-trivial phase encountered is a
homogeneous two-spin superfluid with $\phi_a = \phi_b \ne 0$ ($\phi_\alpha =
\frac{1}{N}\sum_i \phi_{i\alpha}$).
When $n_{a}$ and $n_{b}$ are large enough, $V$ and $U$ terms drive the system
into a 3-fold ordered insulator characterized by $\phi_a = \phi_b=0$ and a
total density $n= 2/3$ $(n_a = n_b=1/3)$, the two-species counterpart of the
$\sqrt{3}\times \sqrt{3}$ phase.
(ii) At even larger density, various supersolids with 3-sublattice structures
are stabilized; they will be discussed in detail in Section \ref{IS}.
Finally, an insulating regime with $n =1$ is stabilized for $U > 1.2V$ in the 
vicinity of $\mu^*=0$; it will be the object of Section~\ref{CS}. For $\mu^* > 0$, 
we obtain an equivalent phase diagram thanks to the particle-hole symmetry.

\section{\label{CS}Commensurate spin-like phases}

\begin{figure}[h!]
\begin{center}
\includegraphics[width=0.4\textwidth]{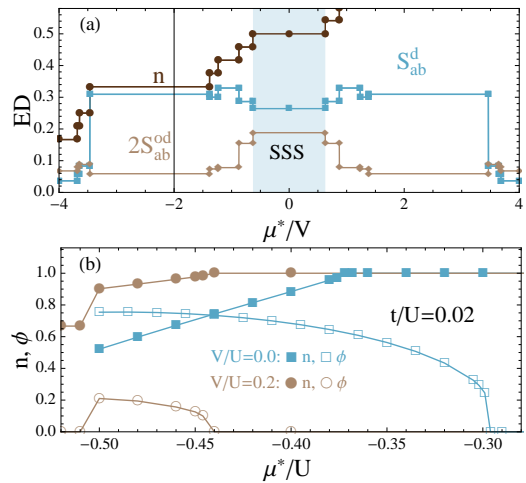}\\
\end{center}
\caption{ \label{EDU2_npV0a}
Strong $U$ regime characterized using either $N=12$ (ED: a) or $N=6$ (CMFT:
b) clusters.
(a) $n$, $S_{ab}^d$ and $S_{ab}^{od}$  as a function of $\mu^*/V$ for $U=2V$
and $t=0.15V$. The $n=1$ plateau and both finite $S_{ab}^d$ and $S_{ab}^{od}$
evidence the SSS; (b) $n$ and $\phi =\phi_a + \phi_b$ as a function of
$\mu^*/U$ distinguishing the SSF ($V=0$) and the SSS ($V=0.2U$) regimes.
}
\end{figure}

For strong inter-species repulsion $U \gg V, t, |\mu^*|$ (triangular uppermost
domain on Fig.~\ref{diag3t015}-up) the system is Mott insulating, with
$\phi_\alpha=0$ and $n=1$. Indeed, $U$ imposes the local constraint of single
occupancy defined as $\bar{n}_{ia} + \bar{n}_{ib} = 1$.
This is reflected by  $n=1$ plateaus in both ED (Fig.~\ref{EDU2_npV0a}-a) and
CMFT (Figs.~\ref{EDU2_npV0a}-b and \ref{scanU05U2b}) results.
In order to better describe this regime, we define spin $1/2$ operators
$\sigma^z_i=(n_{ia}-n_{ib})/2$ and $\sigma^+_i=b^\dagger_{ia}b_{ib}$. At second
order of the perturbation theory, we obtain an effective XXZ model
\begin{eqnarray} H_{\textrm{XXZ}} = -\frac{J_\perp}{2} \sum_{\langle i,j
\rangle} (  \sigma_i^+ \sigma_j^- + \sigma_i^- \sigma_j^+ )  + J_z
\sum_{\langle i,j \rangle} \sigma_i^z \sigma_j^z, \end{eqnarray} where $J_\perp
= 4 t^2/U$ and $J_z = 2V+4t^2/U$.

\begin{figure}[h!]
\begin{center}
\includegraphics[width=0.40\textwidth]{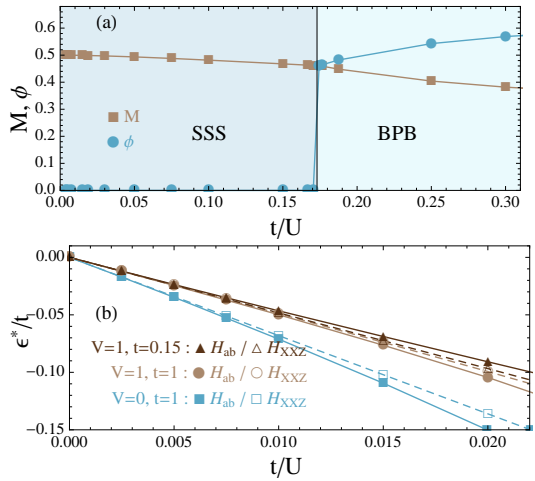}
\end{center}
\caption{ \label{EDU2_npV0b}
Strong $U$ regime characterized using either $N=6$ (CMFT: a) or $N=12$ (ED:
b) clusters.
(a) Species-averaged order parameter $M$ and superfluid fraction $\Phi$, 
evidencing the quantum fluctuation enhancement from the SSS to the BPB phase 
at $V=1$ and $t=0.15$; (b) comparison of the kinetic energy gain 
$\epsilon^*\simeq -k t^2/U$ for $H_{ab}$ (filled) and $H_{\textrm{XXZ}}$ 
(empty) for three ($t,V$) sets.
}
\end{figure}

Interestingly, this model predicts a $(2m_z,-m_z,-m_z)$ SS with diagonal
$\langle \sigma^z\rangle \ne 0$ and off-diagonal $\langle \sigma^+\rangle \ne
0$ order parameters when $J_z/J_\perp > 4.6(1)$; otherwise, a SF with only
off-diagonal order \cite{3XXZ,varXXZ}. In the present context, these phases
correspond respectively to a \textit{spin supersolid} (SSS) and a
\textit{spin superfluid} (SSF).
While the density is uniform, the SSS has a $(2m_z,-m_z,-m_z)$ structure where
$m_z$ quantifies the {\it spin} disproportion on each sublattice.
As depicted in Fig.~\ref{EDU2_npV0a} and Fig.~\ref{EDU2_npV0b}(a), for $|\mu^*| \ll V, U$, 
both ED and CMFT approaches confirm the existence of the SSS; 
indeed the finite structure factors $S_{ab}^{d}$, $S_{ab}^{od}$ and $M = (M_a + M_b)/2$ for
$t/U<0.17$ signal long-range correlations and the corresponding XXZ couplings
verify $J_z/J_\perp = 1+ UV / 2t^2 \geq 20.5$.
In contrast, for $V=0$ and $t/U \le (t/U)_c \simeq 0.05(1)$ we find in vicinity
of $\mu^*=0$ the spatially uniform SSF predicted in the XXZ model, as
illustrated in Fig.~\ref{EDU2_npV0a}(b).
Finally, Fig.~\ref{EDU2_npV0b}(b) shows that the kinetic energy gain
$\epsilon^*\simeq-kt^2/U$ w.r.t. the electrostatic contribution $N(V-\mu)$ is
well reproduced by the XXZ model in both regimes; this validates our approach.

\begin{figure}[h!]
\begin{center}
\includegraphics[width=0.40\textwidth,clip]{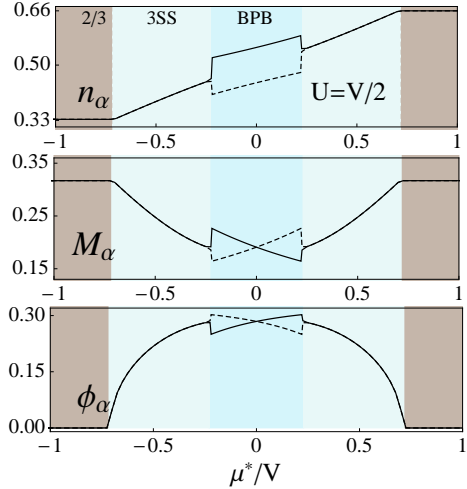}\\
\end{center}
\caption{\label{scanU05U2a}
Species-resolved observables ($n_\alpha, M_\alpha, \phi_\alpha$). Each species
corresponds to either continuous or dashed line.
All data come from a $N=6$ CMFT with fixed $t/V = 0.15$ and $U=0.5V$. 
The arrow points out a tiny region which disappears in the TL (see text).
The shaded regions correspond to the different phases of Fig.~\ref{diag3t015}.
}
\end{figure}

\begin{figure}[h]
\begin{center}
\includegraphics[width=0.40\textwidth,clip]{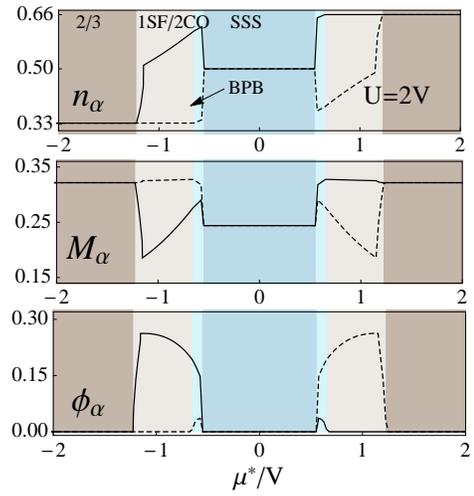}\\
\end{center}
\caption{\label{scanU05U2b}
Same as Fig.~\ref{scanU05U2a}, but for $U = 2V$.
}
\end{figure}
\section{\label{IS} Incommensurate supersolids} 
In contrast to the above mentioned SSS in
which supersolidity comes from the species (spin-like) degrees of freedom, in 
SS phases
discussed here  $n_\alpha$ may be incommensurate and have finite
$\phi_\alpha$.
From the species-resolved observables shown in Fig.~\ref{scanU05U2a} and \ref{scanU05U2b}, we 
identify three such SS phases. 
(i) For large $U/V =2$ (right column) and $\mu^*/V $ about $\pm 1.0(2)$, the
finite $M_\alpha$ indicates a 3-fold order (3FO for both species, while only
one species has a non-zero $\phi_\alpha$. 
This phase, dubbed 1SF/2CO in Fig.~\ref{diag3t015}, is a rearrangement of two
one-species phases of $H_\alpha$, the $\sqrt{3}\times\sqrt{3}$ solid and the SS
obtained from it by enhancing the quantum fluctuations\cite{WesAl,CMF2}.  Two
sublattices are (almost) filled by $a$ and $b$ bosons respectively, while on
the remaining, an incommensurate density for bosons of one species ({\it e.g.}
$a$) accounts for superfluidity ($\phi_a > 0$) and a population imbalance ($n_a
> n_b$).
Within this structure,  the repulsion energy $\propto U$ is minimized thanks to
the localization of bosons of a single species.
(ii) A distinct phase is found for small $U/V =0.5$ (left column) and $\mu^*/V
$ about $\pm 0.5(3)$, with a finite $\phi_\alpha$ for both species as well as a
3-fold symmetry breaking. 

\begin{figure}[h!]
\begin{center}
\includegraphics[width=0.45\textwidth]{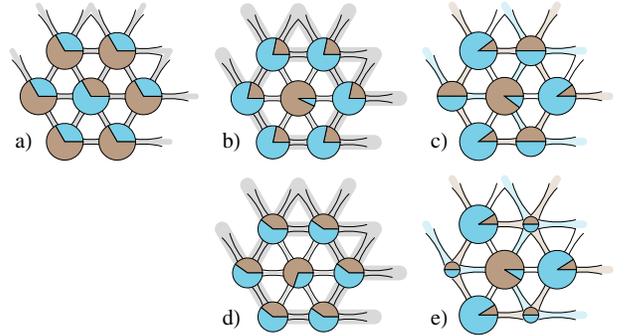}
\caption{\label{snapshots}
Examples of two-spin supersolids with 3-sublattice structure obtained by CMFT
density maps on the 12-site cluster (a,b,c) and by QMC correlations
for $36 \times 36 $ sites cluster (d,e). 
(a) the $n=1$ {\it spin} supersolid (SSS) at $\mu^* =0$ and $U=2V$; (b) and (c)
respectively the bosonic pinball (BPB) at $\mu^*=-0.15V$ and the 3-supersolid
(3SS) at $\mu^* =-0.25V$, for $t=0.15V$ and $U=0.5V$.
$n_a - n_b$ is 0 for the 3SS and finite for the BPB
for all CMFT clusters (Fig.~\ref{scalU05} for the TL study).
(d) and (e) are respectively QMC results for the BPB at $\mu^* = -0.12V$ and the 3SS at
$-0.72V$.
Note that the 3SS and BPB have distinct symmetries (under $\pi/6$ rotations).}
\end{center}
\end{figure}

As shown in the typical snapshot obtained by CMFT in Fig.~\ref{snapshots}(c), it
can be seen as a superposition of two one-species SS (these would be obtained 
at $U=0$); both species contribute symmetrically to  superfluidity.
This phase, called 3-fold SS (3SS) is the first example of a
collective two-species supersolid and is obtained from the parent $n=2/3$ 
solid by a defect condensation upon doping as $\mu^*$ increases.
As in this parent state, the weak inter-species coupling merely forces the
localized $a$ and $b$ bosons to occupy distinct sublattices. 
(iii) The most remarkable incommensurate SS phase is achieved, upon increasing
$t/U$, when quantum fluctuations become too strong for the density-uniform
$n=1$ SSS phase. This original two-species SS is called the {\it bosonic
pinball} (BPB) due to a structure very similar to its fermionic counterpart
with similar interactions, the {\it pinball liquid}, and is depicted in
Fig.~\ref{snapshots}(b,d). The latter has almost one localized electron per site
on one sublattice (pins) and a metallic behavior on the remaining hexagonal
lattice (balls) \cite{Hotta6,CRFMF,MRF}. 
Here, the particles are bosonic and the species play the role of the spins. The
structure is depicted on  Fig.\ref{mopha}(b). One sublattice, forming a
triangular super-lattice with  $\bar{n}_{ia}+\bar{n}_{ib}$ close to 1, is
filled in majority by one type of bosons.  The two-spin superfluid character
is carried by the remaining bosons on the complementary hexagonal lattice.
This is shown in the BPB region on Fig.~\ref{scanU05U2a} and
Fig.~\ref{scalU05}, by a coexistence of solid ($M_\alpha \ne 0$) and superfluid
($\phi_\alpha \ne 0$) orders in both species.
This lattice symmetry breaking is reminiscent of the parent SSS phase, the
partial quantum melting of which involves a condensation of two types of 
defects coming from doubly-occupied and empty sites.
Finally, unlike the 3SS and 1SF/2CO, the BPB can be stabilized at the $n=1$
commensurability when $\mu^* = 0$. 

\section{\label{PT} Thermodynamic limits and phase transitions}
\begin{figure}[!ht]
\begin{center}
\hspace{0.01\textwidth}
\includegraphics[width=0.18\textwidth,clip]{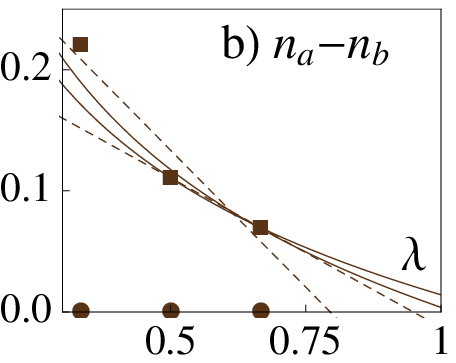}
\hspace{0.01\textwidth}
\includegraphics[width=0.18\textwidth,clip]{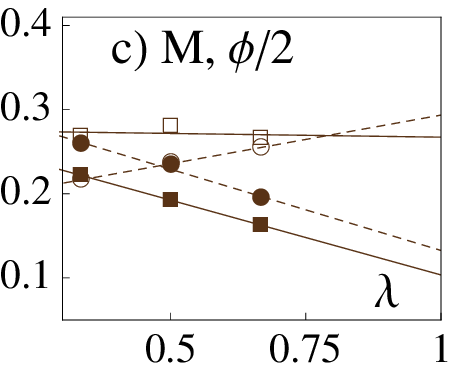}

\includegraphics[width=0.40\textwidth,clip]{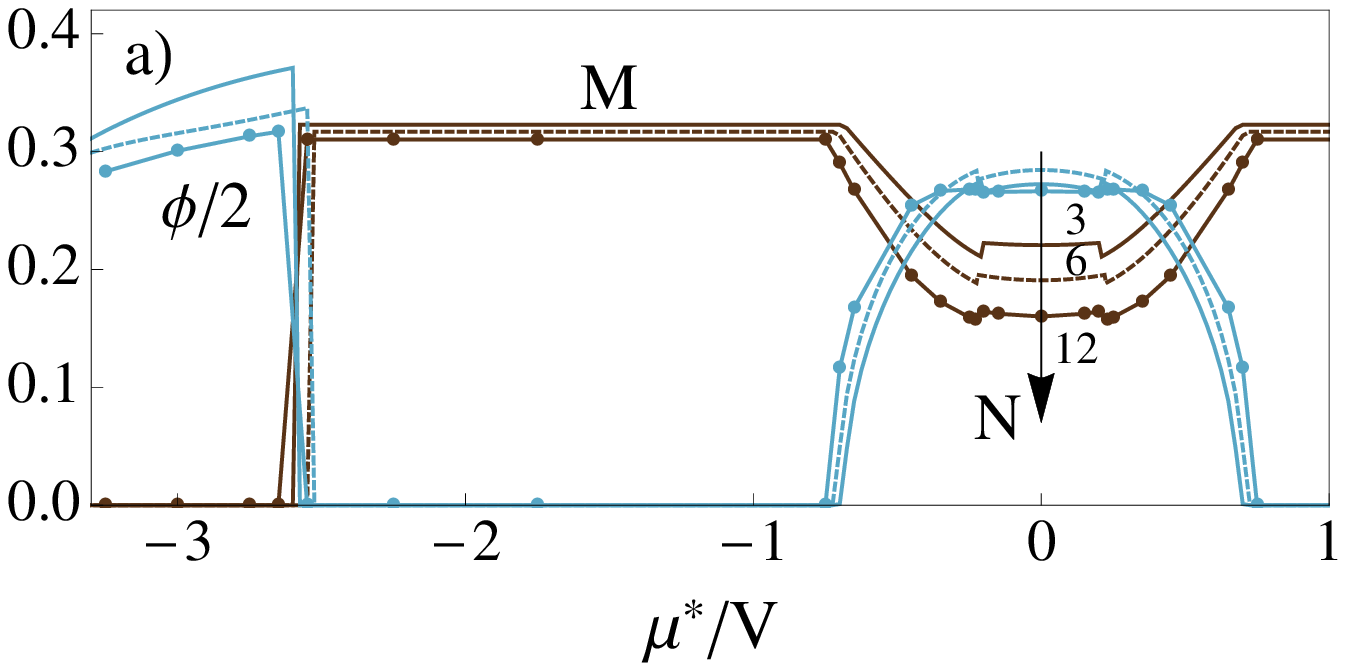}
\caption{ \label{scalU05} $M$ and $\phi$ as a function of $\mu^*$ for $t=0.15V$
and $U=0.5V$ obtained by CMFT on clusters of $N=3$ (continuous), $6$ (dashed)
and $12$ (symbols) site. See Fig.~\ref{clusters} for the cluster shapes.
Cluster scalings as function of $\lambda$ (see text) are shown for b) $n_a -
n_b$ with different possible fits giving either zero (continuous) or finite
(dashed) TL values (see text) and c)  $M$ (filled symbols) and $\phi/2$ (empty
symbols) within both the BPB at $|\mu^*|=t$ (squares) and the 3SS at
$|\mu^*|=3t$ (circles).
}
\end{center}
\end{figure}

All the phases described above exist in the TL. Indeed, in Fig.~\ref{scalU05}
are evidenced the cluster dependencies of $M$ and $\phi$, shown as functions of
$\mu^*$ and  the examples of the BPB and the 3SS (c) prove
their finiteness as $\lambda \to 1$.
We also checked that the phase diagram is only weakly affected by size
effects thus Fig.~\ref{diag3t015} is representative of
the TL.
In the strong correlation limit $U,V\gg t$ [\onlinecite{ZDE}], 
perturbation theory allows to locate phase transitions where the defect 
energy vanishes (condensation), {\it e.g.} for the $2/3 \to$ 3SS transition. 
Consider a defect consisting of an extra boson (say of species $a$) inserted in
the \textit{empty} sublattice of the $2/3$ crystal with the energy cost $3V$.
Via second order processes, such a defect can hop to second neighbors with an
amplitude $-t_{\textrm{eff}}=-t^2/U-t^2/V$. The terms $-t^2/U$ and $-t^2/V$
account respectively for processes where (i) the extra $a$ boson hops via the
$b$-filled sublattice, or (ii) a vacancy on the $a$-filled sublattice is
created and then deleted.
In this limit, the defects condense and lead to supersolidity for $\mu_c=3V-6
t_{\textrm{eff}}$, {\it e.g.} $\mu_c/V \simeq -0.65$ for $U=0.5V$ in good
agreement with the results of Fig.~\ref{scanU05U2a} and \ref{scanU05U2b}.
This defect condensation mechanism implies a gauge symmetry breaking,
additionally to the lattice symmetries already broken in the $2/3$ phase.
This indicates that the transition is of second order, as confirmed by the
absence of discontinuities in $\phi$ and $M$ as function of $\mu$
(Fig.~\ref{scalU05}).
This corresponds to the standard picture of defect condensation accounting for
an incommensurate SS; the SF density (here inhomogeneous) is carried by defects
on top of density modulations \cite{WesAl}.
In contrast, the transitions between (i) homogeneous SF and $2/3$ solid, and
(ii) the BPB and 3SS are found of first order, characterized by hysteresis in
the CMFT. We find distinct transition points if $\mu$ is, from one CMF
calculation to another, either stepwise increased or stepwise decreased (see
[\onlinecite{hyst}]). 
In case (ii), $n_a - n_b$ (Fig.~\ref{scalU05}(b)) has distinct size-dependence,
but a vanishing value in the TL cannot be ruled out for the BPB, despite a
non-linear behavior. $n_a - n_b$ is strictly 0 for the 3SS.  For the BPB, the
non-linear behavior makes difficult to extract $n_a - n_b$ as $\lambda \to 1$,
and various fits can give either zero (linear) or finite ({\it e.g.} weighted
$a + \lambda^b$) value.  However, BPB and 3SS have different symmetries as
obtained by CMFT (Fig.~\ref{snapshots}(b,c)) and confirmed by preliminary QMC
calculations. Indeed, to check this point, we have performed Stochastic Series
Exchange (SSE) QMC simulations on clusters up to $36\times36$ sites for various
$U$ in function of $\mu^*$, for which real-space correlations $\langle n_{sa}
n_{i\alpha}\rangle$(for site $i$ far away from the reference site $s$ to get
rid of short-distance effects) confirm the existence and symmetries of both the
3SS and the BPB (Fig.~\ref{snapshots}(d,e)).
The complete analysis of this model by SSE-QMC, being beyond the scope of this
paper, will constitute a separate work.

\section{\label{CC}Conclusion}
We study an interacting two-species bosonic mixture on a triangular lattice by
combining CMFT, ED and perturbative methods, supported by SSE-QMC
results.
We focus on a region of parameter space in which peculiar phases arise due to the
competition between frustration and quantum fluctuations.
Within a very rich and complex phase diagram, we evidence three original
mixture supersolids, a commensurate {\it spin supersolid} (SSS) and two
mutually competing incommensurate phases arising from the partial quantum
melting of parent states.
The most interesting phase, dubbed {\it bosonic pinball} (BPB) due to an inner
structure very reminiscent of its fermionic counterpart \cite{MRF}, results
from strong inter-species effects. It is worth mentioning that this rich
physics is found for attractive $U$ (for specific parameters), a situation more
directly connected to dipolar cold atom bilayer experiments.
We hope this work will stimulate further investigations in this direction.

\acknowledgements A.R. and F.T. acknowledge financial support by the Agence
Nationale de la Recherche under grant No. ANR 2010 BLANC 0406-0. P.R.-F. would like
to thank N\'eel Institute for kind hospitality.

\end{document}